\documentclass[a4paper]{jpconf}

\usepackage[bookmarks,pdfhighlight=/O,colorlinks=false,pdfstartview=FitH]{hyperref}
\usepackage{amsmath,amssymb,mathrsfs,slashed,bm}
\usepackage{color}
\usepackage{graphicx}
\usepackage{epstopdf}

\newcommand{\MeV}{\mathrm{MeV}}
\newcommand{\GeV}{\mathrm{GeV}}
\newcommand{\TeV}{\mathrm{TeV}}

\begin{document}
\title{Heavy quark transport at RHIC and LHC}

\author{Thomas Lang$^{1,2}$, Hendrik van Hees$^{1,2}$, Jan Steinheimer$^{3}$, \\Yu-Peng Yan$^{4,5}$, Marcus Bleicher$^{1,2}$}

\address{
  $^{1}\,$Frankfurt Institute for Advanced Studies
  (FIAS),Ruth-Moufang-Str. 1, 60438 Frankfurt am Main, Germany }
\address{
  $^{2}\,$Institut f\"ur Theoretische Physik, Johann Wolfgang
  Goethe-Universit\"at, Max-von-Laue-Str. 1, 60438 Frankfurt am Main,
  Germany }
\address{
  $^{3}\,$Lawrence Berkeley National Laboratory, 1 Cyclotron Road, Berkeley, CA 94720, USA
}
\address{$^4$ School of Physics, Institute of Science, Suranaree
University of Technology, Nakhon Ratchasima 30000, Thailand}
\address{$^5$ Thailand Center of Excellence in Physics (ThEP),
Commission on Higher Education, Bangkok 10400, Thailand}

%\ead{lang@fias.uni-frankfurt.de}

\begin{abstract}
  We calculate the heavy quark evolution in heavy ion collisions and 
  show results for the elliptic flow $v_2$ as well as the nuclear 
  modification factor $R_{AA}$ at RHIC and LHC energies. For the calculation 
  we implement a Langevin approach for the transport of heavy quarks in
  the UrQMD (hydrodynamics + Boltzmann) hybrid model. 
  As drag and diffusion coefficients we use a Resonance approach for
  elastic heavy-quark scattering and assume a decoupling temperature of
  the charm quarks from the hot medium of $130\, \MeV$. At RHIC energies 
  we use a coalescence approach at the decoupling temperature for the hadronization of the
  heavy quarks to D-mesons and B-mesons and a sub-following decay to heavy flavor electrons 
  using PYTHIA. At LHC we use an additional fragmentation mechanism to account for the 
  higher transverse momenta reached at higher collision energies. 
\end{abstract}

\section{Introduction}

Heavy quarks are an ideal probe for the QGP. They are produced in the
primordial hard collisions of the nuclear reaction and therefore probe
the created medium during its entire evolution process. When the system
cools down they hadronize, and their decay products can finally be
detected. Therefore, heavy-quark observables provide new insights into
the interaction processes within the hot and dense medium. Two of the
most interesting observables are the elliptic flow, $v_2$, and the
nuclear modification factor, $R_{AA}$, of open-heavy-flavor mesons and
their decay products like ``non-photonic'' single electrons.  The
measured large elliptic flow, $v_2$, of open-heavy-flavor mesons and the
``non-photonic single electrons or muons'' from their decay underline
that heavy quarks take part in the collective motion of the bulk medium,
consisting of light quarks and gluons. The nuclear modification factor
shows a large suppression of the open-heavy flavor particles' spectra at
high transverse momenta ($p_T$) compared to the findings in pp
collisions. This also supports a high degree of thermalization of the
heavy quarks with the bulk medium. 

In this letter we explore the medium modification of heavy-flavor $p_T$
spectra, using a hybrid model, consisting of the Ultra-relativistic
Quantum Molecular Dynamics (UrQMD) model
\cite{Bass:1998ca,Bleicher:1999xi} and a full (3+1)-dimensional ideal
hydrodynamical model \cite{Rischke:1995ir,Rischke:1995mt} to simulate
the bulk medium. The heavy-quark propagation in the medium is described
by a relativistic Langevin approach \cite{Rapp:2009my}. 
Similar studies have recently been
performed in a thermal fireball model with a combined
coalescence-fragmentation approach
\cite{vanHees:2007me,vanHees:2007mf,Greco:2007sz,vanHees:2008gj,Rapp:2008fv,
  Rapp:2008qc,Rapp:2009my}, in an ideal hydrodynamics model with a
lattice-QCD EoS \cite{He:2012df,He:2012xz}, in a model from Kolb and
Heinz \cite{Aichelin:2012ww}, in the BAMPS model
\cite{Uphoff:2011ad,Uphoff:2012gb}, the MARTINI model
\cite{Young:2011ug} as well as in further studies and model comparisons
\cite{Moore:2004tg,Vitev:2007jj,Gossiaux:2010yx,Gossiaux:2011ea,Gossiaux:2012th}.

\section{Description of the model}

The UrQMD hybrid model has been developed to combine the advantages of
transport theory and (ideal) fluid dynamics \cite{Petersen:2008dd}. It
uses initial conditions, generated by the UrQMD model
\cite{Bass:1999tu,Dumitru:1999sf}, for a full (3+1) dimensional ideal
fluid dynamical evolution, including the explicit propagation of the
baryon current. After a Cooper-Frye transition back to the transport
description, the freeze out of the system is treated dynamically within
the UrQMD approach. The hybrid model has been successfully applied to
describe particle yields and transverse dynamics from AGS to LHC
energies
\cite{Petersen:2008dd,Steinheimer:2007iy,Steinheimer:2009nn,Petersen:2010cw,Petersen:2011sb}
and is therefore a reliable model for the flowing background medium.

The diffusion of a ``heavy particles'' in a medium consisting of ``light
particles'' can be described with help of a Fokker-Planck equation
\cite{Svetitsky:1987gq,GolamMustafa:1997id,Moore:2004tg,Svetitsky:1987gq,
GolamMustafa:1997id,vanHees:2004gq,vanHees:2005wb,vanHees:2007me,Gossiaux:2008jv,He:2011yi}
as an approximation of the collision term of the corresponding Boltzmann
equation. It can be mapped into an equivalent stochastic Langevin
equation, suitable for numerical simulations. 

The drag and diffusion coefficients for the heavy-quark propagation
within this framework are taken from a Resonance approach \cite{vanHees:2004gq}, 
where the existence of D-mesons and B-mesons in the QGP phase is assumed, 
as well as a $T$-Matrix approach \cite{vanHees:2007me} in which 
quark-antiquark potentials are used for the calculation 
of the coefficients in the QGP. 

The initial production of charm quarks in our approach is based on a
Glauber approach. For the realization of the initial collision dynamics
we use the UrQMD model. We perform a first UrQMD run excluding
interactions between the colliding nuclei and save the nucleon-nucleon
collision space-time coordinates.  These coordinates are used in a
second, full UrQMD run as possible production coordinates for the charm
quarks.

As momentum distribution for the initially produced charm quarks at
$\sqrt {s_{NN}}=200\; \GeV$ we use
\begin{equation}
\frac{1}{2\pi p_Tdp_T}=\frac{\left(A_1+p_T^2\right)^2}{\left(1+A_2\cdot p_T^2\right)^{A_3}},
\end{equation}
with $A_1=0.5$, $A_2=0.1471$ and $A_3=21$ and for bottom
quarks
\begin{equation}
\frac{1}{2\pi p_Tdp_T}=\frac{1}{\left( A_1+p_T^2 \right)^{A_2}},
\end{equation}
with $A_1=57.74$ and $A_2=5.04$.  These distributions are
taken from \cite{vanHees:2005wb,vanHees:2007me}. 
The $p_T$ distribution for charm quarks at $2.76\,$TeV is 
obtained from a fit to PYTHIA calculations. 
\begin{equation}
\frac{1}{2\pi p_Tdp_T}=\frac{1}{(1+A_1\cdot \left(p_T^2\right)^{A_2})^{A_3}}
\end{equation}
with the coefficients $A_1=0.136$, $A_2=\,2.055$ and $A_3=\,2.862$.

Starting with these distributions as initial conditions we propagate 
the heavy quarks at each hydro-timestep. We use the UrQMD/hydro's cell
velocities, the cell temperature, the size of the time-step, and the
$\gamma$-factor for the calculation of the momentum transfer,
propagating all quarks independently. Our approach provides us only with
the heavy-quark distributions. Since heavy quarks cannot be measured
directly in experiments we include a hadronization mechanism for
D-mesons and B-mesons, via the use of a quark-coalescence mechanism. To implement
this coalescence we perform our Langevin calculation until the
decoupling temperature is reached. Subsequently we add the momenta of
light quarks to those of the heavy quarks.

\section{Results}

\begin{figure}[h]
\begin{minipage}[b]{0.45\textwidth}
\includegraphics[width=1\textwidth]{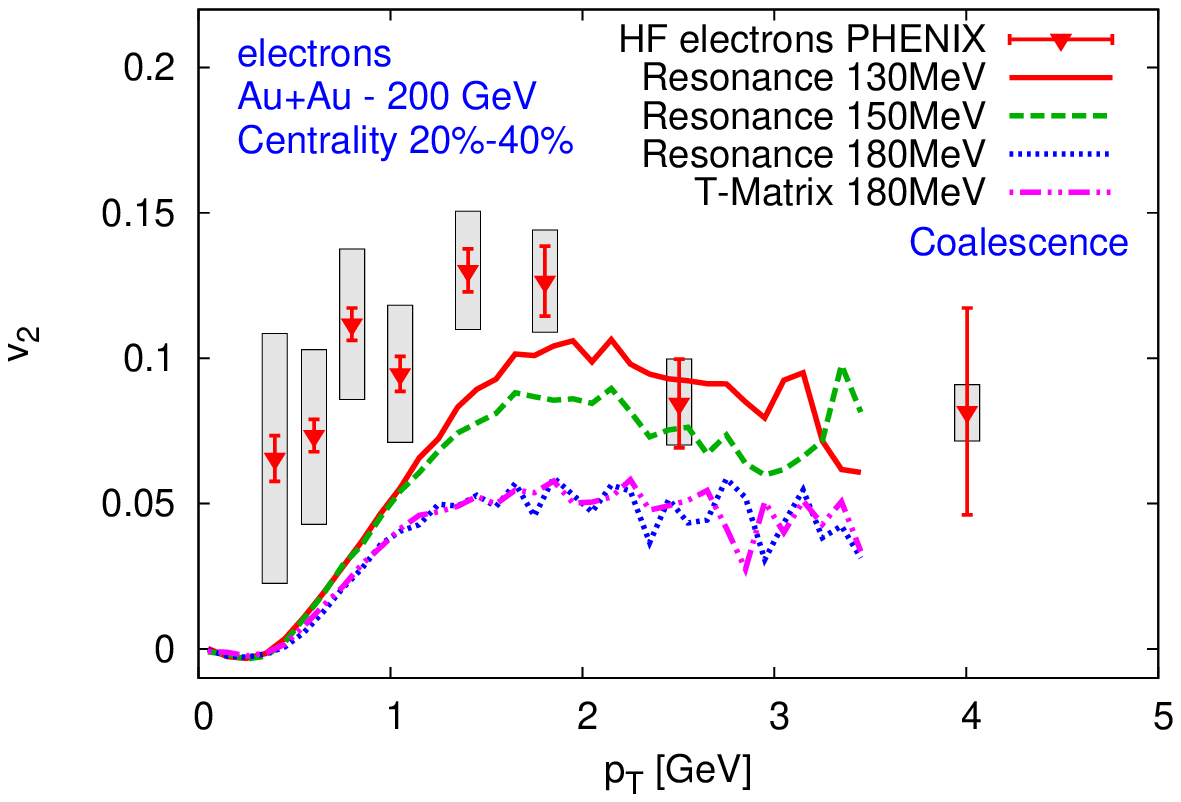}
\end{minipage}
\begin{minipage}[b]{0.45\textwidth}
\includegraphics[width=1\textwidth]{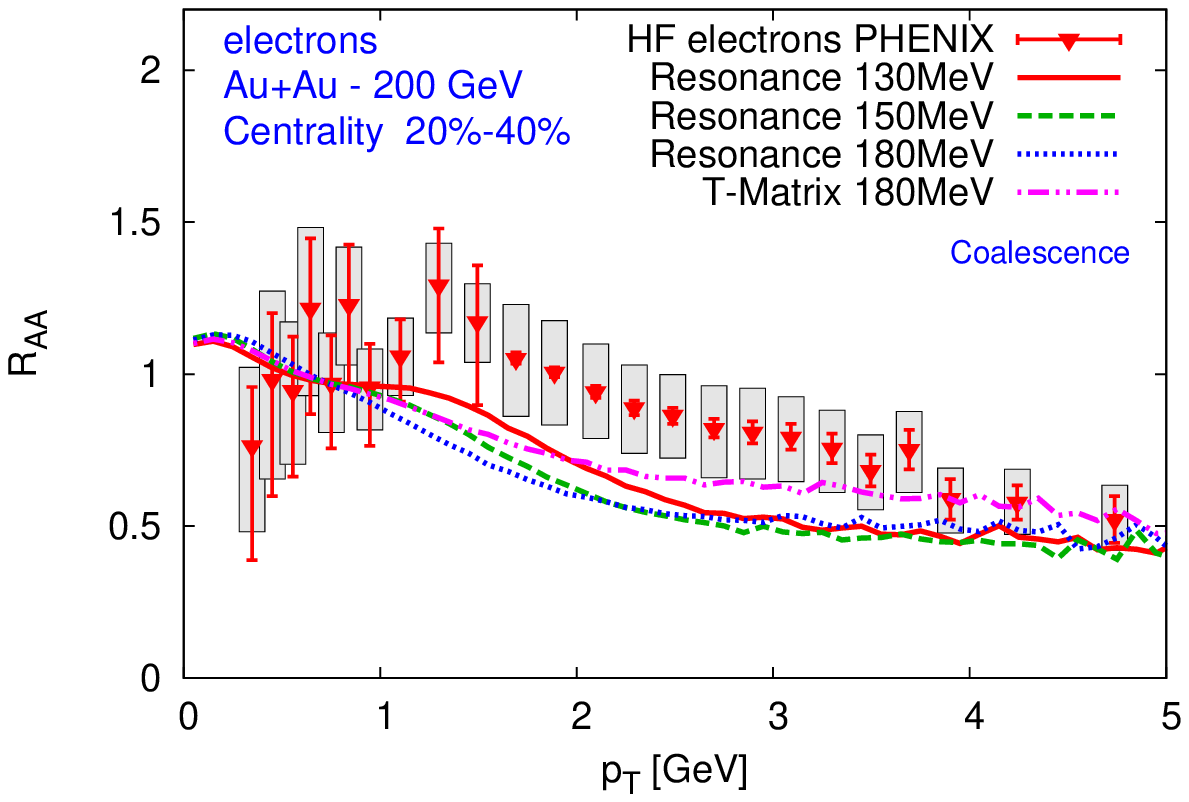}
\end{minipage}
\caption{(Color online) Elliptic flow $v_2$ (left) and nuclear modification factor 
$R_{AA}$ (right) of electrons from heavy quark decays
  in Au+Au collisions at $\sqrt {s_{NN}}=200\;\GeV$ using a coalescence
  mechanism.  We use a rapidity cut of $|y|<0.35$. For a decoupling
  temperature of $130\;\MeV$ we get a reasonable agreement to data
  \cite{Adare:2010de}}
\label{RHIC}
\end{figure}

First we performed our calculations in Au+Au collisions at $\sqrt {s_{NN}}=200\; \GeV$ 
in a centrality range of 20\%-40\%. 
To compare our results to the single-electron spectra measured by PHENIX we use  
PYTHIA for the decay of the heavy quarks to heavy flavor electrons 
and apply a rapidity cut of $|y|<0.35$. 
Fig.~\ref{RHIC} (left) shows our results for the elliptic flow 
$v_2$. For a decoupling temperature of $130\;\MeV$ we obtain a
reasonable agreement with the experimental data except for low $p_T$ bins. 
Here a depletion effect can be seen. This effect is due to the radial
velocity of the medium, which is in case of a developed elliptic flow
larger in $x$ than in $y$ direction. Consequently there is a depletion
of particles with high $v_x$ in the low $p_T$ region and smaller
elliptic flow. This effect is more important for heavier particles and a
larger radial flow \cite{Huovinen:2001cy,Krieg:2007bc}.

In Fig. \ref{RHIC} (right) the nuclear modification factor $R_{AA}$ for non-photonic 
single electrons is depicted.

Also here we obtain a good agreement with the data, especially in case of using 
the T-Matrix coefficients or a low decoupling temperature. 

\begin{figure}[h]
\begin{minipage}[b]{0.45\textwidth}
\includegraphics[width=1\textwidth]{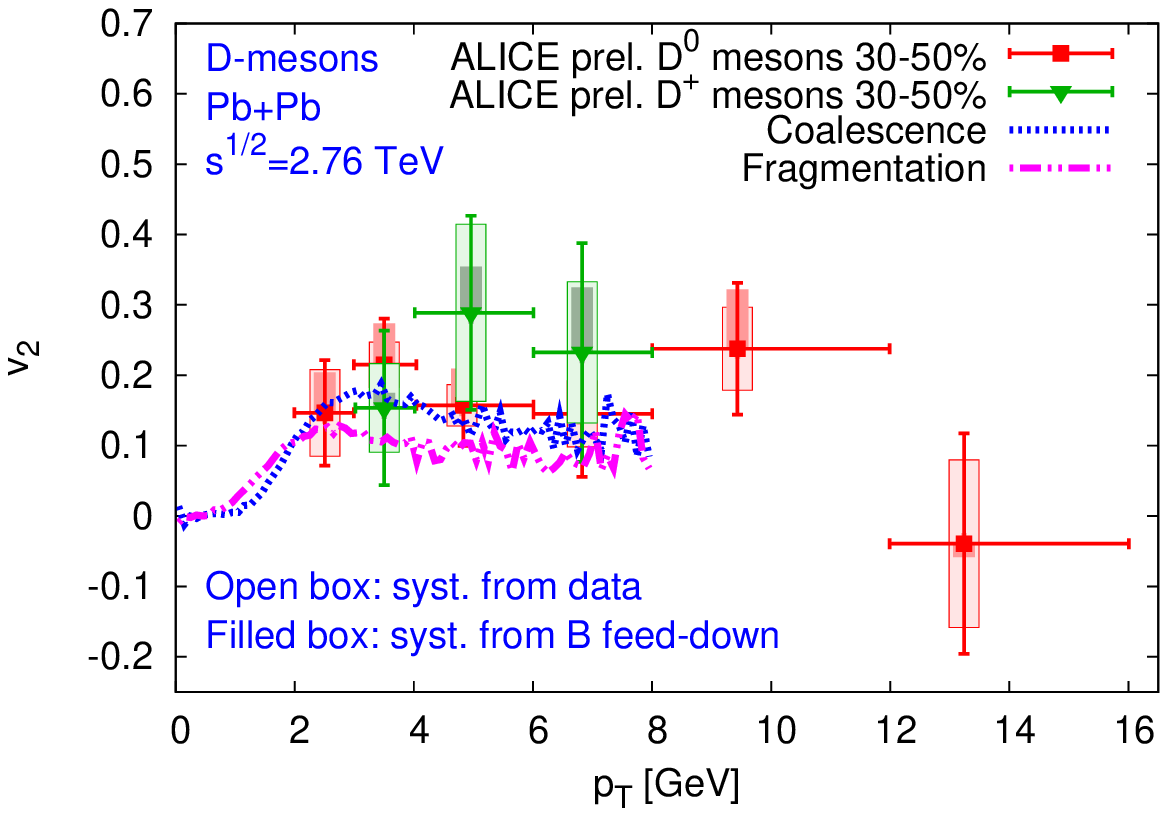}
\end{minipage}
\begin{minipage}[b]{0.45\textwidth}
\includegraphics[width=1\textwidth]{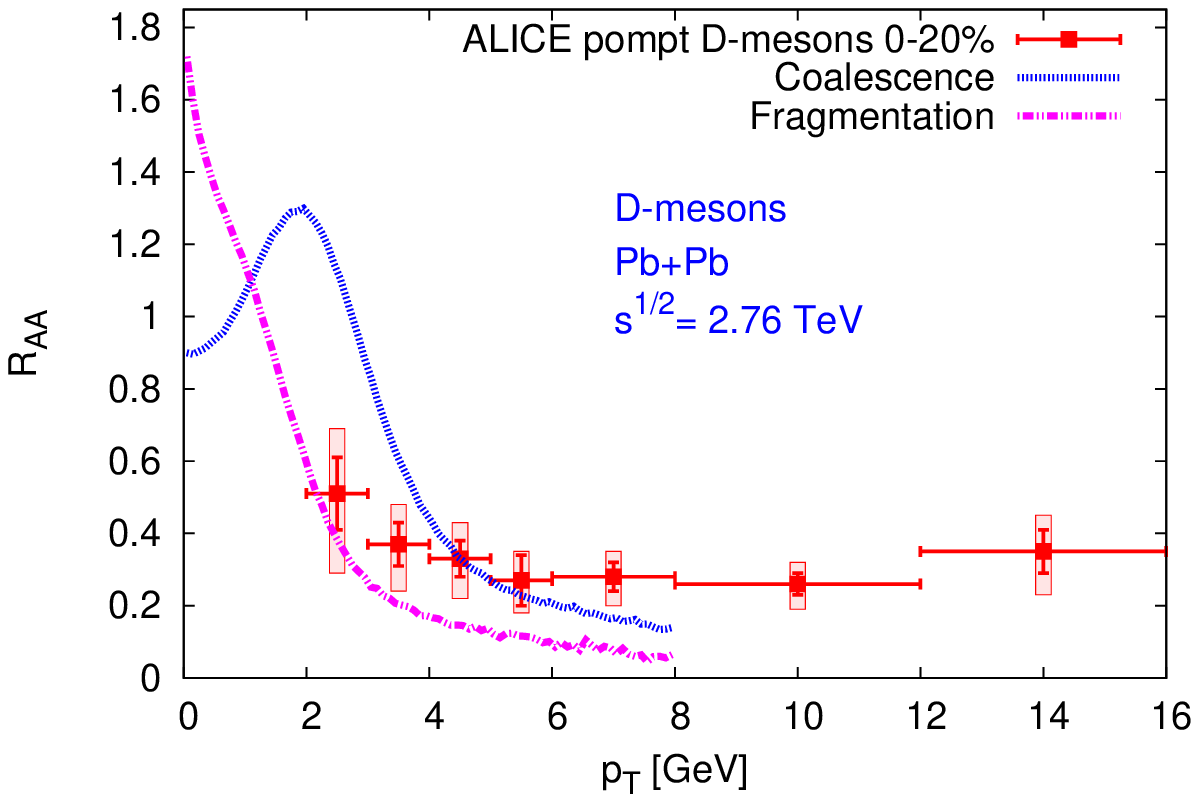}
\end{minipage}
\caption{(Color online) Left: Flow $v_2$ of D-mesons in Pb+Pb collisions at $\sqrt
  {s_{NN}}=2.76\,$TeV compared to data from the ALICE experiment. 
(Talk by Z.~Conesa del Valle at QM 2012, data not published yet.) 
A rapidity cut of $|y|<0.35$ is employed. Right: $R_{AA}$ of D-mesons in Pb+Pb collisions at $\sqrt
  {s_{NN}}=2.76\; \TeV$ compared to experimental data from ALICE
  \cite{ALICE:2012ab}. A rapidity cut of $|y|<0.35$ is employed.}
\label{LHC}
\end{figure}

Now we performed the same calculations, but in Pb+Pb collisions at $\sqrt{s}_{NN}
=2.76\; \TeV$ in a centrality range of 30\%-50\%. The analysis is done
in a rapidity cut of $|y|<0.35$ in line with the ALICE data. 
Here we made use of the coalescence mechanism with a decoupling temperature of $130\,\text{MeV}$ only since 
we achieved the best results using this configuration at RHIC energies.  
In the ALICE experiment D-mesons are measured. Therefore we do not need to perform the 
decay to electrons this time. 
Fig.\ \ref{LHC} (left) depicts our results for the elliptic flow compared
to ALICE measurements.
Additionally, apart form the calculation using the coalescence mechanism, also 
a calculation using a fragmentation mechanism is shown, since fragmentation 
might get more important at higher $p_T$ bins, as measured at LHC. 
As fragmentation mechanism we used the Peterson fragmentation \cite{Peterson:1982ak}. 
$$D^H_Q(z)=\frac{N}{z[1-(1/z)-\epsilon_Q/(1-z)]^2},$$
Here $N$ is a normalization constant, $z$ the relative-momentum
fraction obtained in the fragmentation of the charm quarks and
$\epsilon_Q=0.05$. 

Both $v_2$ calculations are in agreement with the ALICE data set. 
Using the fragmentation function a sharper rise of the elliptic flow at low $p_T$ is reached, 
while at medium $p_T$ the flow using the coalescence approach is stronger. 
At high $p_T$ both hadronization mechanisms lead to similar results. 

A complementary view on the drag and diffusion coefficients is provided
by the nuclear suppression factor $R_{AA}$.  Figure \ref{LHC} (right) shows
the calculated nuclear modification factor $R_{AA}$ of D-mesons at
LHC. Here we compare to two data sets available, for $D^0$ and $D^+$
mesons. In line with the experimental data the simulation is done for a
more central bin of $\sigma/\sigma_ {to}=0\%$-$20\%$. 
In case of the coalescence approach we find a maximum of the $R_{AA}$ at about $2 \; \GeV$ followed by a
sharp decline to an $R_{AA}$ of about $0.2$ at high $p_T$. 
The fragmentation approach leads to a different result at low $p_T$. 
A very sharp $R_{AA}$ drop-off from low to high $p_T$ is seen. 
At high $p_T$ the two approaches nearly converge. 
Concerning the difference of the results using the fragmentation and coalescence mechanism  
new $v_2$ and $R_{AA}$ measurements, especially at low $p_T$, would be very helpful to draw conclusions 
on the hadronization mechanism at LHC.

To summarize, we presented in this letter our results on the medium modification of heavy quarks at RHIC and LHC energies 
using the nuclear modification factor $R_{AA}$ and the elliptic flow $v_2$ as observables. 
At RHIC energies we compared different sets for drag and diffusion coefficients and obtained the best agreement to 
experimental measurements if using a Resonance model with a decoupling temperature of $130\,\text{MeV}$. 
At LHC we compared a coalescence approach and a fragmentation approach as hadronization mechanism using the Resonance model 
at a decoupling temperature of $130\,\text{MeV}$. Both approaches describe the ellitpic flow $v_2$ in pretty good agreement 
with the experimental data while for the $R_{AA}$ a major disagreement between our models at low $p_T$ can be seen 
that needs to be resolved by new measurements.

\section{ACKNOWLEDGMENTS}

We are grateful to the Center for Scientific Computing (CSC) at
Frankfurt for providing computing resources. T.~Lang gratefully
acknowledges support from the Helmholtz Research School on Quark Matter
Studies.  This work is supported by the Hessian LOEWE initiative through
the Helmholtz International Center for FAIR (HIC for
FAIR). J.~S. acknowledges a Feodor Lynen fellowship of the Alexander von
Humboldt foundation.  This work is supported by the Office of Nuclear
Physics in the US Department of Energy's Office of Science under
Contract No. DE-AC02-05CH11231.

\end{document}